\documentclass[usenatbib,onecolumn]{mn2e}
\usepackage{psfig}

\title[Fowler-Nordheim Electron Cold Emission Formalism in Presence of Strong Magnetic Field]
{Fowler-Nordheim Electron Cold Emission Formalism in Presence of Strong Magnetic Field}
\author
[Arpita Ghosh and Somenath Chakrabarty]{Arpita Ghosh and Somenath
Chakrabarty\thanks{E-mail: somenath.chakrabarty@visva-bharati.ac.in}\\
Department of Physics, Visva-Bharati, Santiniketan 731235, India}
\begin{document}

\date{Accepted: To be put by the Editor ; Received: ; 
in original form:To be put by the
Editor}

\pagerange{\pageref{firstpage}--\pageref{lastpage}} \pubyear{2011}

\maketitle

\label{firstpage}

\begin{abstract}
Formalisms for both non-relativistic as well as relativistic versions of field emission
of electrons in presence of strong quantizing magnetic field, relevant for strongly
magnetized neutron stars or magnetars are developed. In the non-relativistic scenario,
where electrons obey Schr{$\ddot{\rm{o}}$}dinger equation, we have noticed that when 
Landau levels are populated for electrons in presence of strong quantizing magnetic field the 
transmission probability exactly vanishes unless the electrons are spin polarized in the
opposite direction to the external magnetic field. On the other hand,
the cold electron emission under the influence of strong
electrostatic field at the poles is totally forbidden from the
surface of those compact objects for which the surface magnetic field
strength is $\gg 10^{15}$G (in the eventuality that they may exist).
Whereas in the relativistic case, 
where the electrons obey Dirac equation, the presence of strong quantizing magnetic field
completely forbids the emission of electrons from the surface of  
compact objects if $B >10^{13}$G.
\end{abstract}


\section{Introduction}

There are mainly three kinds of electron emission
processes from metal surface, they are: (i) thermionic emission, (ii)
photoelectric emission and (iii) cold emission or field emission.

The field emission or cold emission, which we have investigated in the present article in the 
context of
strongly magnetized neutron stars or magnetars, 
is an electron emission process induced by strong external
electrostatic field at zero or 
extremely low temperature. Field emission can happen from solid and liquid 
surfaces, or from individual atoms. 
It has been noticed that the field emission from metals occurs in presence of high 
electric field: the gradients are typically higher than 1000 volts per micron 
and the emission is strongly dependent upon the work function of the material.
Unlike the thermionic emission and photo-emission of electrons,
the field emission process can only be explained by quantum 
tunneling of electrons, which has no counter classical explanation. However, 
for general type surface barrier, this purely quantum mechanical problem
can not be solved exactly, a semi-classical approximation, known as 
WKB (the name is an acronym for Wentzel-Kramers-Brilloun) is needed to get tunneling coefficients. Now to explain cold electron emission from
metals, one may assume that because of quantum fluctuation,
electrons from the sea of conduction electrons (degenerate electron gas) always try to 
tunnel out through the metallic surface (surface barrier). However, as soon as an electron comes 
out, it induces an image charge on the metal surface, which pulls it back
and does not allow this emitted electron to move far away from metal surface in the atomic scale. 
But if some strong attractive electrostatic field is applied near the metallic surface, 
then depending on the Fermi energy of electrons, the height of the surface barrier 
and the local work function, the electrons may overcome the
effect of image charge and get liberated. Since the external
strong electric field is causing such emission and does not depend on
the thermal properties of the metal, even the metal can be at zero
temperature, it is called field emission or cold emission.

The theory of field emission from bulk metals was first proposed by
Fowler and Nordheim in an epoch making paper in the proceedings of
Royal Society of London in the year 1928 \cite{FN1} (see also
\cite{FN2,FN3,FN4,nano} for further discussion). 
Fowler-Nordheim tunneling is the wave-mechanical tunneling of electrons
through a triangular type barrier produced at the surface of an
electron conductor by applying a very high electric field.  

Now the cold emission or field emission processes not only have significance in
the terrestrial laboratories, but is found to be equally important in the
electron emission processes from cold and compact stellar objects, such as
neutron stars. 
In the case of a rotating neutron star, 
the existing large magnetic field, $\geq 10^{12}$G \cite{ST} for the conventional
radio pulsars or $\geq 10^{15}$G for the magnetars inner field \cite{mag},
produces a strong electric field 
at the poles, approximately given by $F\sim 2\times
10^8P^{-1}B_{12}$ Volt cm$^{-1}$  and is parallel to $\vec B$ at the poles \cite{ST},
here $P$ is the time period of the neutron star in second and $B_{12}$ is the measure of
magnetic field strength in the units of $10^{12}$.

At the proximity of polar region of a strongly magnetized neutron star, the 
potential difference changes almost
linearly with distance from the polar cap, which is a region very
close to the magnetic pole. The repulsive surface barrier 
in combination with this attractive potential, forms a triangular type 
barrier at the poles. Therefore if electron emission from the poles of neutron
stars is field emission type, then Fowler-Nordheim equation with
proper modification may be used to investigate such emission process. Now,
the study of plasma formation in a pulsar magneto-sphere is a quite old 
but still an unresolved astrophysical issue, in particular the magneto-spheres of strongly 
magnetized neutron stars/magnetars \cite{VM1,VM2,YD}. In the formation of 
magneto-spheric plasma, it is generally assumed that there must be an initial high energy 
electron flux from the magnetized neutron stars. Since the magnetic field
at the poles of neutron 
stars/magnetars are strong enough, the emitted electrons flow only along the magnetic 
field lines. The flow of high energy electrons along the direction of magnetic 
lines of forces and their penetration through the light cylinder is conventionally pictured 
with the 
current carrying conductors. Naturally, if the conductor is broken near the 
pulsar surface the entire potential difference will be developed across a thin 
gap, called polar gap. This is of course based on the assumption that
above a critical height from the polar gap, because of high electrical 
conductivity of the plasma, the electric field $F$, parallel to the 
magnetic field near the poles is quenched. Further, a steady acceleration of
electrons 
originating at the polar region of neutron stars, travelling along the field 
lines, 
will produce magnetically convertible curvature $\gamma$-rays.
If these curvature $\gamma$-ray photons have energies $>2m_ec^2$ (with $m_e$ is
the electron rest mass and $c$ is the velocity of light), then pairs of $e^--e^+$ 
will be produced in enormous amount with very high efficiency near the polar 
gap. These produced $e^--e^+$ pairs form what is known as the magneto-spheric 
plasma \cite{ST,R2,R3,R4,R6,R66,R7,R8}. The cold emission, therefore plays a
significant role in magneto-spheric plasma formation. In turn, the
motion of charged particles in the magnetosphere in presence of strong magnetic field causes pulsar
emission in the form of synchrotron radiation. Therefore the cold emission process
indirectly also affects the intensity of synchrotron radiation.

Further the exactly solvable models with simple type tunneling barrier lead to equations
\cite{FN1,FN2} that underestimates the emission current density by a factor 
of 1000 or more. If a more realistic type barrier model is used by inserting 
an exact surface potential in the simplest form of the 
Schr${\ddot{\rm{o}}}$dinger equation, then a complicated mathematical 
problem arises over the resulting differential equation.
It is in principle therefore mathematically impossible to solve the
equation exactly in terms of the usual functions of mathematical physics, 
or in any simple way. 

Moreover, to the best of our knowledge, neither the non-relativistic nor the relativistic
version of cold emission processes in presence 
of strong quantizing magnetic field, relevant for electron emission
from the poles of strongly magnetized neutron stars/magnetars, even with
simple type potential barriers have been properly investigated.
In the conventional pulsar model it is generally assumed that 
the emission of electrons and thereby formation of magnetosphere 
is mainly caused by strong electric field at the polar region which is produced 
by the strong magnetic field of rotating neutron stars. Taking 
this physical picture into consideration, in this article, we have developed formalisms
for both non-relativistic and relativistic scenarios of field emissions for electrons from
the poles of neutron stars with $10^{10}{\rm{G}}\leq B \leq 10^{17}$G.

In the next section we have studied the effect of strong quantizing magnetic field
on the field emission of electrons for the non-relativistic case. 
In section-3, we have repeated the same calculation for the relativistic scenario.
In the last section we have given the conclusions and future prospects of this work.
\section{Effect of Strong Quantizing Magnetic Field on Cold Emission: Non-relativistic
Scenario}
To develop the modified version of field emission process for the electrons in the non-relativistic
scenario in presence of a strong quantizing magnetic field, we have followed the basic
calculation presented in the seminal Royal Society paper by \cite{FN1}. In the modified version, 
we assume a cylindrical co-ordinate system
$(\rho,\theta,z)$ and the constant magnetic field $\vec B$ is along positive $z$-direction, with the 
usual gauge for the vector potential $\vec A=(\vec B \times \vec \rho)$. Following \cite{FN1}, we
assume that the triangular shape surface potential is given by $V(z)=C-Fz$, which is
changing linearly with $z$-coordinate. Here $C$ is the constant surface barrier and $F$
(absorbing the magnitude of electron charge $e$, we replace $eF$ by $F$) is the driving field for
electron emission from the poles. 
Consideration of linear type surface barrier potential has
no-doubt some historical importance. This type of potential barrier
was first used in the original work by Fowler and Nordheim. However,
there are two other important reasons: since this is the first time the problem
is solved in presence of strong quantizing magnetic field, therefore
to get an analytical solution, we have considered such simplest triangular
type potential barrier. Our intension is to solve the problem
analytically in a way in which the physical meaning of the problem is not 
lost. The other 
reason behind such a choice is
that, in the case of strongly magnetized rotating neutron stars /
magnetars, the produced electric field at the polar region is approximately
constant for uniform magnetic field strength at that region and
constant rotational period of the object. We believe that for the
potential barrier at the poles both assumptions are approximately
valid. Then assuming the
conservative force field relation
$dV/dz=-eF$ we get the triangular type potential barrier at the poles. Under
such situation, the Schr{$\ddot{\rm{o}}$}dinger equation satisfied by the electrons which
are confined within the matter (in this case within the neutron star/magnetar crustal
matter) is
given by (throughout the paper for the sake of convenience we assume natural units, i.e., 
$\hbar=c=1$)
\begin{equation}
-\frac{1}{2m}\left [\frac{1}{\rho}\frac{\partial}{\partial \rho}\left ( \rho \frac{\partial
\psi}{\partial \rho } \right )+\frac{1}{\rho^2}\frac{\partial^2\psi}{\partial \theta^2}
+\frac{\partial^2\psi}{\partial z^2} \right ]-\frac{ieB}{2m} \frac{\partial \psi}{\partial
\theta} +\left (\frac{e^2B^2\rho^2}{8m_e}-E\right )\psi =0
\end{equation}
Whereas for the electrons just liberated out through quantum mechanical tunneling, one has
to consider the potential $V(z)$ along with $E$. The relevant equation is given by
$$
-\frac{1}{2m}\left [\frac{1}{\rho}\frac{\partial}{\partial \rho}\left ( \rho \frac{\partial
\psi}{\partial \rho } \right )+\frac{1}{\rho^2}\frac{\partial^2\psi}{\partial \theta^2}
+\frac{\partial^2\psi}{\partial z^2} \right ]-\frac{ieB}{2m} \frac{\partial \psi}{\partial
\theta} +\left (\frac{e^2B^2\rho^2}{8m_e}-E+V(z)\right )\psi =0 \eqno(1a)
$$
where the energy eigen value $E$ is given by the eqns.(4) and (5) for two different physical
situations. If we assume a separable solution for the eqns.(1) and (1a), 
satisfied by freely moving electrons and moving under the potential $V(z)$, the wave
functions can be represented by
\begin{equation}
\psi(\rho,\theta,z)=\phi_{n_\rho,m}(\rho,\theta)f_\nu(z)
\end{equation}
where for eqn.(1), the longitudinal part is plane wave type, whereas for eqn.(1a), we shall
evaluate $f_\nu(z)$ using the technique as discussed below. 
Since there is no potential associated with the transverse motion for the electrons, the free 
transverse part of the wave function is given by
\begin{eqnarray}
\phi_{n_\rho,m}(\rho,\theta)&=&\frac{\exp(im\theta)}{(2\pi)^{1/2}}\rho_0^{-1-\mid m\mid} \left [ \frac{(\mid
m \mid +n_\rho )!}{2^{\mid m \mid} n_\rho!\mid m \mid !}\right ] \nonumber \\ &\times&
\rho^{\mid m \mid} \exp\left (-\frac{\rho^2}{4\rho_0^2}\right )L_{n_\rho,m} \left(
\frac{\rho^2}{2\rho_0^2} \right )
\end{eqnarray} 
with $\rho_0=(2/eB)^{1/2}$ is the Larmor radius, the radius of the lowest Landau orbit and
$L_{n_\rho,m}$ is the Laguerre polynomial. The energy eigen value is given by 
\begin{equation}
E=\frac{p_z^2}{2m_e}+\mu_B B(2n_\rho+\nu\pm\nu+1)
\end{equation}
without the electron spin and with the inclusion of the electron spin, it will be
\begin{equation}
E=\frac{p_z^2}{2m_e}+\mu_B B(2n_\rho+\nu\pm\nu+1)\mp\mu_B B
\end{equation}
where $\mu_B=e/2m$, the Bohr magneton.
For the motion of electrons along $z$-direction, the Schr{$\ddot{\rm{o}}$}dinger equation 
satisfied by
$f_\nu(z)$ with and without potential $V(z)$ can be obtained by averaging eqns.(1) and (1a)
respectively over the transverse wave
function $\phi_{n_\rho,m}(\rho,\theta)$. Then for just tunneled out electrons in the lowest
Landau level with 
$n_\rho=\nu=m=0$, moving along $z$-axis under the influence of the potential $V(z)$ and
having no spin contribution, we have
\begin{equation}
-\frac{1}{2m}\frac{d^2f_0}{dz^2}+V(z)f_0=\left ( E-\mu_B B \right ) f_0 = w f_0
\end{equation}
Whereas for the same kind of liberated electrons with spin polarization in the negative
direction of $z$-axis, we have $w=E$ and for the polarization along the direction of
magnetic field, $w=E-2\mu_BB$.
On the other hand, for the electrons confined within
the crustal matter of magnetars and moving freely, we have $V(z)$=0. 
The corresponding Schr{$\ddot{\rm{o}}$}dinger equation is given by 
\begin{equation}
\frac{d^2f_0}{dz^2}+w_k^2f_0=0
\end{equation}
Where $w_k=(2mw)^{1/2}$, is the equivalent electron momentum along $z$-axis and $w$ can
have three possible values as mentioned above. 
Then following eqn.(10) of \cite{FN1}, we can write down the solution of eqn.(7) for free
electrons inside the crustal matter in the form
\begin{equation}
f_0=\frac{1}{w_k^{1/2}}\left [a\exp(iw_kz)+a^\prime\exp(-iw_kz)\right]
\end{equation}
where $a$ is the probability amplitude for electrons moving along the 
positive direction of $z$-axis (incident part),
whereas $a^\prime$ is the corresponding quantity for left moving
waves (reflected part from the surface barrier). 
Next for the electrons just tunneled out, we make the following coordinate transformation
\begin{equation}
y=\left (-\frac{C-w}{F}+z\right ) (2m_eF)^{1/3}
\end{equation}
Then we have from eqn.(6)
\begin{equation}
\frac{d^2f_0}{dy^2}+yf_0=0
\end{equation} 
Now following eqn.(8) of \cite{FN1} the solution for this equation is given by 
\begin{equation}
f_0(y)=y^{1/2}H_{1/3}^{(2)}\left (\frac{2}{3}y^{3/2} \right )
\end{equation}
where $H_{1/3}^{(2)}(x)$ is the Hankel function of second kind of order $1/3$ with 
argument $x$.
As we shall see below  
the quantity $Q$ defined in \cite{FN1} after eqn.(12) is much larger in our formalism in
presence of strong quantizing magnetic field, unless the direction of spin polarization for the
emitted electrons are opposite to the direction of external magnetic field.
In absence of electron spin term or polarization along the direction of magnetic field, the factor
$Q$ is virtually infinitely large ($Q\approx \infty$). We shall further show that this
infinitely large $Q$ will make the transmission probability of electrons vanishingly small.
In our formalism the factor $Q$ is defined as
$$
Q=\frac{2}{3}(2m_eF)^{1/2}\left ( \frac{C-w}{F}\right )^{3/2}\eqno(11a)
$$
and is related to the Hankel function argument by the relation
$$
H^{(2)}_{1/3}\left [\exp\left(-\frac{3}{2}\pi i\right )Q\right ]
~~{\rm{at}}~~ z=0\eqno(11b).
$$
Now the transmission probability for electrons as defined in \cite{FN1} is given by  
\begin{equation}
D(w)=\frac{\mid a\mid^2-\mid a^\prime \mid^2}{\mid a \mid^2}
\end{equation}
Since the transmission coefficient $D$ related to $Q$ factor by the relation $D\sim
\exp(-2Q)$, and the field emission current for the electrons from the zeroth Landau level is
related to $D$ by the integral
$$
R=\frac{eB}{2\pi^2}\int_0^\infty f(w)D(w)\frac{p_z}{m_e}dp_z\eqno(12a)
$$
with $f(w)$ the Fermi distribution function, then it is quite obvious that in the electron
field emission current, the polarization
effect will come only from the transmission coefficient $D(w)$. In
the following, we therefore focus our study in the investigation of
the properties of the emission coefficient.
For large $Q$, we have from eqn.(18) of \cite{FN1} in our modified form
\begin{eqnarray}
D(w)&\approx& \frac{[w(C-w)]^{1/2}}{C}\exp\left ( -\frac{4}{3}(2m_eF)^{1/2}
\left (\frac{C-w}{F}\right )^{3/2}
\right )\nonumber \\ &=&\frac{[w(C-w)^{1/2}]}{C}\exp\left( -2Q\right )
\end{eqnarray}
Let us now analyze the argument part of the exponential. Following
\cite{FN1}, we put $C=\mu_e+w_f$, where $\mu_e$ is the electron Fermi energy and
$w_f=w_c\times (B/B_c^{(e)})^{1/2}$ in eV is the work function (see \cite{jaa}) for the emission 
of electrons along
the direction of magnetic field, where $w_c\approx 82.93$ and $B_c^{(e)}\approx 4.43\times
10^{13}$G, the typical value of magnetic field strength at which the
Landau levels for the electrons are populated in the relativistic
scenario. Again, as defined before, the quantity $w$ can have three possible values. To get an 
order of 
magnitude estimate for the terms containing Bohr magneton and work function, we assume that
the emitted electrons carry the maximum possible  
energy, i.e., the electron Fermi energy for temperature $T\longrightarrow 0$ limit. 
Then $C-w=w_f+\beta\times \mu_B B$, where the parameter $\beta=0$, or $=1$ or
$=2$ for the spin polarization opposite to the direction of external magnetic field, i.e.,
with conventional direction of spin polarization in presence of strong magnetic field, or no spin 
term or spin
polarization along the direction of magnetic field respectively. Taking into account 
the denominator $F$ of the 
argument and expressing in terms of magnetic field strength as defined at the beginning of the
introduction and using $\mu_B\approx 5.79 \times 10^{-15}$MeV
G$^{-1}$, the numerical value for the Bohr magneton, we have approximately from the expression
for $Q$ as defined above
\begin{equation}
Q\approx h_f^{1/2}(\beta \times 0.5+w_c h_f^{-1/2}\times 10^{-6})^{3/2}\times 10^7=Q_B+Q_{wf}
\end{equation}
where $h_f=B/B_c^{(e)}$, $Q_B$ and $Q_{wf}$ are the contributions from Bohr magneton or
spin term and
work function part respectively. It is quite obvious that the contribution from spin term 
$Q_B$ is extremely large for $\beta=1$ or $\beta=2$ and
makes the transmission coefficient exactly zero. Which physically means that if we do not
consider electron spin or assume electron spin polarization along the direction of magnetic field,
the electron transmission coefficient and in turn the electron transmission current
vanishes exactly, i.e., there will be no field emission under such situations. On the other hand, 
the second term, the work
function part, unlike the first term, gives finite contribution to transmission coefficient. 
The first term,
i.e., the spin term will make $Q$ extremely high, even if we do not assume that the
emitted electrons are at the top of their Fermi level. This particular term is therefore
making the cold emission
probability of electrons from the poles of strongly magnetized neutron stars exactly zero
for the non-zero values of $\beta$. For the transmission of spin polarized electrons
$\beta=0$ and consequently $Q_B=0$. Therefore $Q=Q_{wf}$. In this situation the electron energy
eigen value obtained from eqn.(5) for $n_\rho=\nu=m=0$ is given by
\begin{equation}
E=\frac{p_z^2}{2m_e}
\end{equation}
In this case there will be electron field emission with their spins polarized opposite to
the direction of external magnetic field.
From the above expressions for electron energy eigen value, it is quite obvious that for
our model on cold emission of polarized electrons 
the rest of the
mathematical formulation will be almost identical with that of Fowler and
Nordheim in presence of strong quantizing magnetic field. We have further noticed that the
values for transmission coefficient $D$ and the transmission current $R$ remain
finite and large enough for the magnetic field strength $\leq
10^{15}$G.
Of course in our model the magnetic field 
dependency of $Q$ will come from the work function $w_f$ and the field intensity $F$ and
only the polarized electrons are allowed to tunnel through the surface barrier.
Then following Fowler and Nordheim we have obtained the cold electron emission current from
eqn.(12a) at $T\longrightarrow 0$ using Sommerfeld's lemma. While
calculating emission 
current numerically, we have used the exact form
of $D$ as given in \cite{FN1} after eqn.(16). In fig.(1) we have shown schematically the
variation of cold electron current with the strength of electric field intensity at the
poles and is represented by the solid curve.  
Since the electric field at the poles is produced by the rotating magnetic field, we have
also shown in the same graph with dashed curve the variation of electron cold current  with the 
intensity of magnetic field. Since the electric field intensity varies linearly with the
magnetic field strength, the qualitative nature of the curves are almost identical. 
For the neutron stars with very low surface magnetic field, the produced electric field,
which acts as driving force, is also
small enough, therefore the electron field emission current will be extremely small as
shown in the curve. For the neutron stars with moderate surface
magnetic field strength ($B\sim 10^{12}-10^{15}$G) the
field current is quite high. Now for the objects with large surface
magnetic field ($B\gg 10^{15}$G), the
work function will also become large. Therefor beyond some maximum value for electron
field current, since the work function part dominates over the driving electric force at
the poles, the electron field current will decrease and will become
vanishingly small. This is the case for
the objects with ultra-high surface magnetic field ($\gg 10^{15}$G). In the figure, we have not shown the
variation of electron field current with the magnetic field strength
(computed at the surface) for a
particular neutron star. Therefore each magnetic field / electric field points corresponds a
particular type
of compact magnetized object, with surface magnetic field from very low to ultra-high
values. From the curves it is quite obvious that for magnetars with
surface magnetic field $\sim 10^{15}$G, the electron field current is quite high and very
close to the peak value. Unlike the original work of 
Fowler and Nordheim the tunneling coefficient does not follow exponential law. However, the 
variation of cold current with the electric field strength can be obtained
numerically. The numerically fitted functional form is given by 
\begin{equation}
R=0.26 F_{24}^{1/2} \exp(-9.8 \times F_{14})
\end{equation}
where $R$ is the field current in Amp/cm$^2$, $F_{14}=10^{-14}F$ and $F_{24}=10^{-24}F$. 
\section{Effect of Strong Quantizing Magnetic Field on Cold Emission: Relativistic
Scenario}
In the relativistic scenario we have repeated the non-relativistic calculation for the 
cold emission transmission co-efficient. In this section we have considered emission of high 
energy electrons from the polar region of strongly magnetized neutron
stars with magnetic field $10^{12}\leq B \leq 10^{17}$ in Gauss.
The potential is introduced by hand in the Dirac equation
using standard relativistic hadro-dynamic technique. Following Fowler and Nordheim, here
also we have considered a triangular type potential barrier at the polar region.
Like the previous case, we have considered cylindrical coordinate system and the choice of gauge
is same for $\vec A$. The radial part of upper component satisfies the equation
\begin{equation}
\left [ \beta_\lambda+\frac{\partial^2}{\partial \rho^2} +\frac{1}{\rho}
\frac{\partial}{\partial \rho}-k^2\rho^2\right ]R(\rho)=0
\end{equation}
where $\beta_\lambda^2=E^2-{m^*}^2-p_z^2+2\lambda k$ with $\lambda=\pm 1$ for up and down
spin states respectively, $k=eB/2$, $E$ is the energy eigen value for
the Dirac equation and $m^*=m+V(z)=m+C-Fz$ is the effective electron mass and $m^*=m$ for free
electrons when $V(z)=0$. The solution of the above equation is given by
\begin{equation}
R(\rho)=N\exp\left ( -\frac{t}{2}\right ) L_n(t)
\end{equation}
where $L_n(t)$ is the Legendre polynomial of order $n$, $t=k^2\rho^2$ and $N$ is the normalization 
constant. For up and down spin states, the energy eigen values are given by
$E_\uparrow=[p_z^2+m^2+2neB]^{1/2}$ and $E_\downarrow=[p_z^2+m^2+2(n+1)eB]^{1/2}$
respectively. In general we
may write $E_\nu=(p_z^2+m^2+2\nu eB)^{1/2}$. Then following the same
averaging technique as discussed
in the previous section for the non-relativistic case, the wave function of 
electrons
corresponding to the motion along $z$-direction is given by
\begin{equation}
\frac{d^2f_\nu}{dz^2}+(E^2-{m^*}^2-2\nu eB)f_\nu=0
\end{equation}
Using the transformation 
\begin{equation}
z=\frac{m+C-uF^{1/2}}{F}
\end{equation}
with the new variable being $u$, the above equation for $f_\nu$ reduces to
\begin{equation}
\frac{d^2f_\nu}{du^2}+(\alpha^2-u^2)f_\nu=0
\end{equation}
where $\alpha=(E^2-2\nu eB)^{1/2}/F^{1/2}$. 
which is the well known form of differential equation for one dimensional quantum 
mechanical harmonic oscillator. 
With $\alpha^2=2l$, we have $E^2=2(\nu eB+lF)$, and the solution is given by
\begin{equation}
f_{\nu,l}=i\tilde{N}\exp\left (-\frac{u^2}{2}\right) H_l(u)=\left ({f_\nu}
\right )_{I}~~{\rm{(say)}}
\end{equation}
where $\tilde{N}$ is the normalization constant and $H_l(u)$ is the
Hermite polynomial of order $l$. This spinor solution $f_{\nu,l}$ is for
those electrons which have already been liberated out through the
surface into vacuum under the influence of electric field $F$ (here
liberated out from the crustal matter of strongly magnetized neutron stars or magnetars into the
magnetosphere through polar region). 

The equation satisfied by free electrons bound within the crustal matter by the barrier
potential at the surface can be obtained by putting $V(z)=0$ and 
is given by
\begin{equation}
\frac{d^2f_\nu}{dz^2}+\alpha^2f_\nu=0
\end{equation}
where in this case, $\alpha=(E^2-m^2-2\nu e B)^{1/2}$ is the free electron
momentum along $z$-axis of energy $E$. Now following the 
notation of \cite{FN1}, 
we express the solution for free electrons within the system, 
confined by the surface barrier $V(z)$, in the form
\begin{equation}
f_\nu=\frac{1}{\alpha^{1/2}}\left [a\exp(i\alpha z)+a^\prime\exp(-i\alpha z)\right]=
\left (f_\nu \right)_{II}~~{\rm{(say)}}
\end{equation}
where as before $a$ is the probability amplitude for electrons moving along the 
positive direction of $z$-axis (incident part),
whereas $a^\prime$ is the corresponding quantity for left moving
waves (reflected part from the surface barrier). 
Assuming the interface between the crustal matter of the strongly magnetized
neutron stars and the magnetosphere is 
at $z=0$, the wave
function and their derivatives must be continuous at $z=0$ \cite{FN1}, i.e.,
\begin{equation}
\left (f_\nu(0)\right )_{II}=\left (f_\nu(0)\right
)_{I}~~~{\rm{and}}~~~\left (f_\nu^\prime(0)\right )_{II}=\left
(f_\nu^\prime(0)\right )_{I}
\end{equation}
Using the relation $H_l^\prime(u)=2\nu H_{l-1}(u)$, we have 
\begin{equation}
a+a^\prime=N \alpha^{1/2}\exp\left
(-\frac{u_0^2}{2}\right)H_l(u_0)
\end{equation}
and
\begin{equation}
i\alpha^{1/2}(a-a^\prime)=N\exp\left (-\frac{u_0^2}{2}\right)\left [u_0
H_l(u_0)-2l H_{l-1}(u_0)\right]
\end{equation}
where $u_0=u(z=0)=m^*/F^{1/2}$. These two conditions may be rearranged in the 
following form
\begin{equation}
a+a^\prime=X ~~{\rm{and}}~~ a-a^\prime=iY
\end{equation}
where $X$ and $Y$ are the two real quantities. Hence it is straight forward to
verify from eqn.(12) that in the relativistic scenario the transmission coefficient  
vanishes exactly. Therefore from the analysis of this section, we may conclude that if the barrier 
in combination with the external electrostatic driving force behaves like a 
scalar type potential and is triangular in shape at the surface, then the relativistic
electrons can not tunnel through the surface barrier whatever be their kinetic energies and the 
strength of external electric field.
\section{Conclusion}
The non-relativistic scenario of cold electron emission in presence of strong magnetic
field is believed to be the first attempt in this direction.
While obtaining the electron transmission probability in the non-relativistic scenario under the
influence of strong electric field at the poles, we have noticed that in our theoretical
formalism, the emission is allowed if we take electron spin into
account and also the electrons have conventional spin polarization, i.e.,
opposite to the direction of external magnetic field. Even if all these criteria are
satisfied, at extremely high magnetic field strength since the electron
work function becomes large enough, the transmission coefficient drops
to zero. Since there are no such stellar objects with
surface magnetic field strength $\gg 10^{15}$G, the vanishingly small
transmission coefficients will therefore not be possible in reality.
The low charge density magneto-sphere will therefore only be possible
if such super exotic compact stellar objects with ultra-strong
surface magnetic field exist in nature. In the conventional magnetar
or strongly magnetized neutron star case, the electron field current is quite high, very
close to the peak value, which
is also obvious from the figure. As a result there will be enough
curvature gamma photons produced by the energetic electrons, which in turn produce 
enormous amount of
$e^--e^+$ pairs to form normal magneto-spheric plasma. In our model the
only difference from non-magnetic or low magnetic emission case is
that the primary electrons are spin polarized in the direction
opposite to the direction of external magnetic field. As a consequence, for the
conventional magnetar case, the results obtain in \cite{APJ} on the formation of
corona of magnetars will not be seriously affected.
Therefore, in the magnetar magnetosphere of a neutron star with $B
\leq 10^{15}$G, the
primary electrons are all spin polarized along $-\vec B$. Since the electron emission
current is almost zero for the objects with ultra-high magnetic field strength ($\gg
10^{15}$G),
then if the electron emission process in such exotic objects is dominated by cold emission, 
the charge density of $e^--e^+$-plasma in the magnetosphere will be extremely
low. As a consequence there will be very weak synchrotron emission in the 
radio wave band. The other possible mechanism by which $e^--e^+$-plasma can be produced in the magnetosphere of
magnetars are (i) thermoelectric emission of electrons from the polar region and (ii) photo 
emission from the same region. However the work function at the polar region of a typical 
magnetar is several GeV, whereas the temperature can be at most a few hundred MeV for an young 
magnetar. The thermionic emission will therefore be suppressed by the Boltzmann factor 
$\exp(-w_f/kT)$. Whereas in the case of photo emission, the energy of the induced photon must 
be of GeV order ($\gamma$-photons). At the polar region, if any such photons exist to liberate 
electrons, they must have produced as curvature photons by high energy electrons moving along the 
magnetic lines of forces. Since the possibility of such
electrons is very rare, the number of high frequency photon is also vanishingly small, as a 
result there will be almost no creation of secondary $e^--e^+$-pairs in the photo emission 
process. 

In conclusions, from our relativistic formalism of cold emission of
electrons we can state that relativistic electrons populating the
neutron star interior can not be extracted from cold emission from
the poles of a neutron star, independently from the magnetic field
strength. Non-relativistic electrons with anti-parallel spin can be
extracted for standard (observed) values of magnetic field strengths,
but can not be extracted from the surface of objects with $B\gg
10^{15}$G (in the eventuality that such exotic objects can exist).

\newpage
\begin{figure}
\psfig{figure=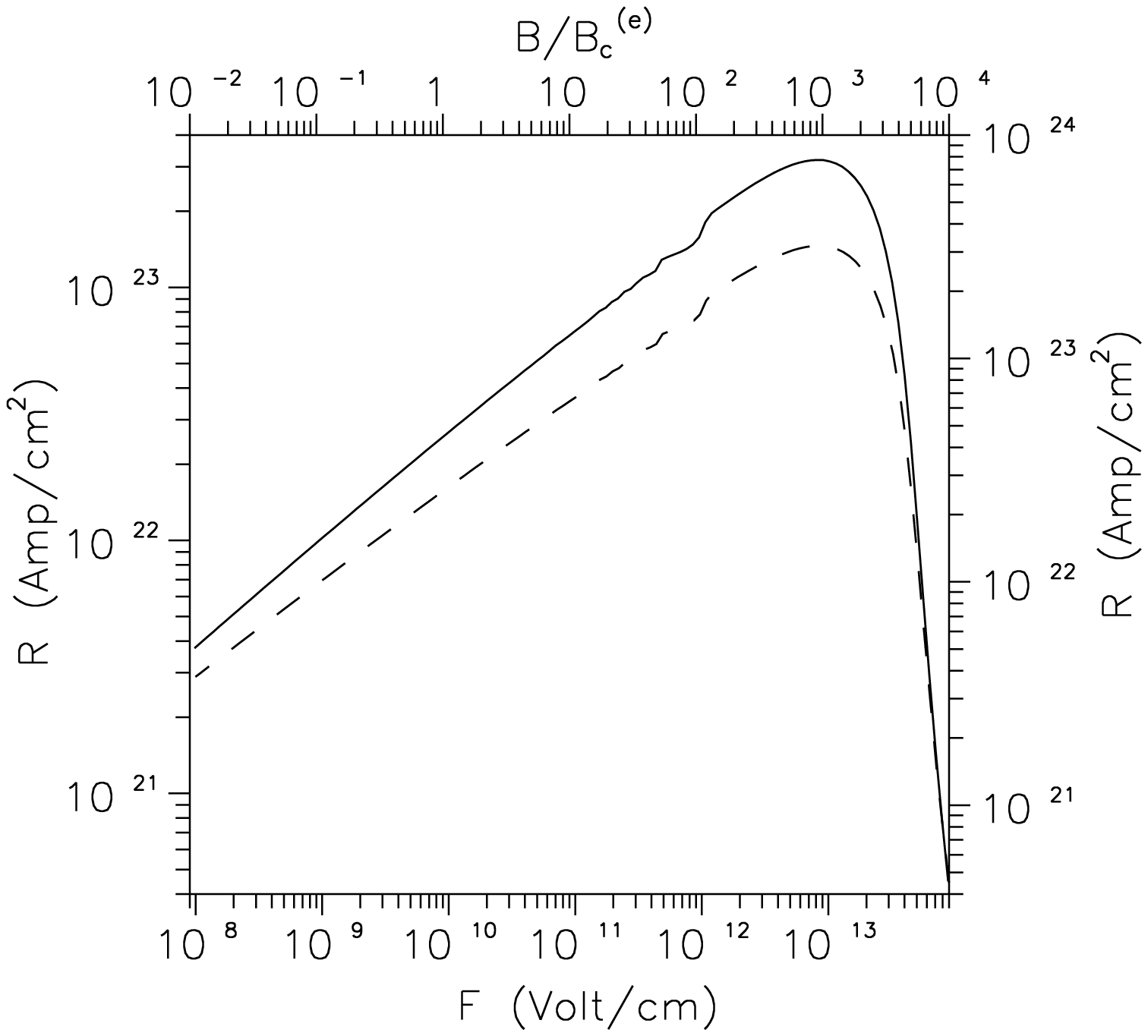,height=0.5\linewidth}
\caption{(i) With dashed curve the variation of electron field current plotted along left side 
vertical axis with the neutron star magnetic field
plotted along $x$-axis at the bottom and expressed in terms of critical field 
$B_c^{(e)}$ for electron is shown. (ii) The variation of same quantity plotted along right side 
vertical axis with the electric field in Volt/cm produced by rotating magnetic
field of magnetars, plotted along horizontal axis at the top is shown by
solid curve.}
\end{figure}
\bsp
\label{lastpage}

\begin{thebibliography}{}
%
\bibitem[Beloborodov et al. (2007)]{APJ} Beloborodov A.M. and Thompson C., 2007, APJ, 657,
967.
\bibitem[Diver et al. (2009)]{R4} Diver D.A., da Costa A.A., Laing E.W., Stark C.R. and 
Teodoro L.F.A., 2009, astro-ph/0909.3581.
\bibitem[Forbes \& Deane (2007)]{FN4} Forbes R.G and Deane J.H.B, 2007, 
Proc. Roy. Soc. London, 463, 2907.
\bibitem[Fowler \& Nordheim (1928)]{FN1} Fowler R.H,
Nordheim Dr. L., 1928, Proc. R. Soc. London 119, 173.
\bibitem[Ghosh \& Chakrabarty (2011)]{jaa} Ghosh A. and Chakrabarty S., 2011, J.
Astrophys.  Astr., 32 377. 
\bibitem[Harding \& Lai (2006)]{R7} Harding A.K. and Lai D., 2006, Rep. Prog. Phys., 69, 2631.
\bibitem[Istomin \& Sobyanin (2007)]{YD} Istomin Ya.N. and Sobyanin D.N., 2007, arXiv:0710.1000.
\bibitem[Jensen (1995)]{FN3} Jensen K.L, 1995, J. Vac. Sci. Technol., B13,  516.
\bibitem[Jessner et al. (2001)]{R2} Jessner A., Lesch H. and Kunzl T., 2001, APJ, 547, 959. 
\bibitem[Liang \& Chen (2008)]{nano} Liang Shi-Dong and Chen Lu, 2008, Phys. Rev. Lett., 101, 027602.
\bibitem[Mereghetti (2009)]{mag} Mereghetti S., 2009, astro-ph/0904.4880.
\bibitem[Michel (1982)]{R6} Michel F.C., 1982, Rev. Mod. Phys., 54, 1.
\bibitem[Michel (2004)]{R66} Michel F.C., 2004 Advances in Space Research, 33, 542.
\bibitem[Molofeev et al. (2004)]{VM1} Molofeev V.M, Malov O.I. and D. A. Teplykh
D.A., 2004, in Proceedings of the IAU Symposium No.
218: Young Neutron Stars and Their Environments, Ed. dy F. Camilo and B. M. Gaensler (Astron. Soc.
Pac., San Francisco), p.261
\bibitem[Molofeev et al. (2005)]{VM2} Molofeev V.M, Malov O.I. and Teplykh D.A., 2005, Astron. 
Rep. 49, 242.
\bibitem[Ruderman (1971)]{R8} Ruderman M., 1971, Phys. Rev. Lett., 27,  1306.  
\bibitem[Ruderman \& Sutherland (1975)]{R3} Ruderman M.A. and Sutherland P.G., 1975, APJ, 196, 51.
\bibitem[Shapiro \& Teukolsky (1983)]{ST} Shapiro S.L. and Teukolsky S.A., 1983, Black Holes, 
White Dwarfs,and Neutron Stars- The Physics of Compact Objects, John Wiley \& Sons,
New York, 288.
\bibitem[Stern et al. (1929)]{FN2} Stern T.E, Gossling B.S and Fowler R.N, 1929, Proc. 
Roy. Soc. London, 124, 699.
\end{thebibliography}
\end{document}